\pdfoutput=1
\documentclass[12pt,a4paper]{article}

\usepackage{ifthen} 
\newboolean{pdflatex}
\setboolean{pdflatex}{true} 

\newboolean{articletitles}
\setboolean{articletitles}{true} 

\newboolean{uprightparticles}
\setboolean{uprightparticles}{false} 

\def\paperauthors{LHCb collaboration} 
\def\paperasciititle{Precision measurement of the muon charge asymmetry from W-boson decays in pp collisions at sqrt(s) = 13 TeV with the LHCb detector} 
\def\papertitle{Precision measurement of the muon charge asymmetry from $W$-boson decays in $pp$ collisions at\\
$\sqrt{s}$ = 13\tev in the forward region} 
\def\paperkeywords{{High Energy Physics}, {LHCb}} 
\def\papercopyright{\the\year\ CERN for the benefit of the LHCb collaboration} 
\def\paperlicence{CC BY 4.0 licence}
\def\paperlicenceurl{https://creativecommons.org/licenses/by/4.0/}

\def\WToMuNu     {\decay{W}{\mu \nu}}

\newif\ifEnableSectionTOCLinks
\EnableSectionTOCLinksfalse 

\usepackage[top=1in, bottom=1.25in, left=1in, right=1in]{geometry}

\usepackage{microtype}
\usepackage{lineno}  
\usepackage{xspace} 
\usepackage{caption} 

\usepackage{graphicx}  
\usepackage{color}
\usepackage{colortbl}
\graphicspath{{./figs/}} 

\usepackage{amsmath} 
\usepackage{amssymb}
\usepackage{amsfonts}
\usepackage{upgreek} 

\newcommand*\patchAmsMathEnvironmentForLineno[1]{%
\expandafter\let\csname old#1\expandafter\endcsname\csname #1\endcsname
\expandafter\let\csname oldend#1\expandafter\endcsname\csname
end#1\endcsname
 \renewenvironment{#1}%
   {\linenomath\csname old#1\endcsname}%
   {\csname oldend#1\endcsname\endlinenomath}%
}
\newcommand*\patchBothAmsMathEnvironmentsForLineno[1]{%
  \patchAmsMathEnvironmentForLineno{#1}%
  \patchAmsMathEnvironmentForLineno{#1*}%
}
\AtBeginDocument{%
\patchBothAmsMathEnvironmentsForLineno{equation}%
\patchBothAmsMathEnvironmentsForLineno{align}%
\patchBothAmsMathEnvironmentsForLineno{flalign}%
\patchBothAmsMathEnvironmentsForLineno{alignat}%
\patchBothAmsMathEnvironmentsForLineno{gather}%
\patchBothAmsMathEnvironmentsForLineno{multline}%
\patchBothAmsMathEnvironmentsForLineno{eqnarray}%
}

\usepackage[pdftex,
            pdfauthor={\paperauthors},
            pdftitle={\paperasciititle},
            pdfkeywords={\paperkeywords}]{hyperref}
\usepackage{hyperxmp}
\hypersetup{
    pdfcopyright={Copyright (C) \papercopyright},
    pdflicenseurl={\paperlicenceurl}
}

\usepackage[colorinlistoftodos,textsize=scriptsize]{todonotes}

\usepackage[bottom,flushmargin,hang,multiple]{footmisc}

\usepackage[all]{hypcap} 

\usepackage{xspace} 
\usepackage{upgreek}

\def\lhcb   {\mbox{LHCb}\xspace}
\def\atlas  {\mbox{ATLAS}\xspace}
\def\cms    {\mbox{CMS}\xspace}

\def\lhc    {\mbox{LHC}\xspace}

\def\MagUp {\mbox{\em Mag\kern -0.05em Up}\xspace}

\ifthenelse{\boolean{uprightparticles}}%
{

 \def\Pmu         {\ensuremath{\upmu}\xspace}

 \def\Ppsi        {\ensuremath{\uppsi}\xspace}

 \def\PDelta      {\ensuremath{\Delta}\xspace}                 
 \def\PXi         {\ensuremath{\Xi}\xspace}                 
 \def\PLambda     {\ensuremath{\Lambda}\xspace}                 
 \def\PSigma      {\ensuremath{\Sigma}\xspace}                 
 \def\POmega      {\ensuremath{\Omega}\xspace}                 
 \def\PUpsilon    {\ensuremath{\Upsilon}\xspace}
 \let\oldPi\Pi
 \def\PPi         {\ensuremath{\oldPi}\xspace}

 \def\PB      {\ensuremath{\mathrm{B}}\xspace}                 
 \def\PD      {\ensuremath{\mathrm{D}}\xspace}                 
 \def\PJ      {\ensuremath{\mathrm{J}}\xspace}                 
 \def\PK      {\ensuremath{\mathrm{K}}\xspace}                 
 \def\PW      {\ensuremath{\mathrm{W}}\xspace}                 
 \def\PZ      {\ensuremath{\mathrm{Z}}\xspace}                 
 \def\Pb      {\ensuremath{\mathrm{b}}\xspace}                 
 \def\Pc      {\ensuremath{\mathrm{c}}\xspace}                 
 \def\Pd      {\ensuremath{\mathrm{d}}\xspace}                 
 \def\Pe      {\ensuremath{\mathrm{e}}\xspace}                 
 \def\Ps      {\ensuremath{\mathrm{s}}\xspace}                 
 \def\Pt      {\ensuremath{\mathrm{t}}\xspace}                 
 \def\Pu      {\ensuremath{\mathrm{u}}\xspace}                 
 \def\thebaroffset{0.0em}
}
{

 \def\Pmu         {\ensuremath{\mu}\xspace}

 \def\Ppsi        {\ensuremath{\psi}\xspace}                 
                  
 \mathchardef\PDelta="7101
 \mathchardef\PXi="7104
 \mathchardef\PLambda="7103
 \mathchardef\PSigma="7106
 \mathchardef\POmega="710A
 \mathchardef\PUpsilon="7107
 \mathchardef\PPi="7105
 \def\PB      {\ensuremath{B}\xspace}                 
 \def\PD      {\ensuremath{D}\xspace}                 
 \def\PJ      {\ensuremath{J}\xspace}                 
 \def\PK      {\ensuremath{K}\xspace}                 
 \def\PW      {\ensuremath{W}\xspace}                 
 \def\PZ      {\ensuremath{Z}\xspace}                 
 \def\Pb      {\ensuremath{b}\xspace}                 
 \def\Pc      {\ensuremath{c}\xspace}                 
 \def\Pd      {\ensuremath{d}\xspace}                 
 \def\Pe      {\ensuremath{e}\xspace}                 
 \def\Ps      {\ensuremath{s}\xspace}                 
 \def\Pt      {\ensuremath{t}\xspace}                 
 \def\Pu      {\ensuremath{u}\xspace}                 
 \def\thebaroffset{0.18em}
}
\newcommand{\offsetoverline}[2][\thebaroffset]{\kern #1\overline{\kern -#1 #2}}%

\makeatletter
\ifcase \@ptsize \relax
  \newcommand{\miniscule}{\@setfontsize\miniscule{4}{5}}
\or
  \newcommand{\miniscule}{\@setfontsize\miniscule{5}{6}}
\or
  \newcommand{\miniscule}{\@setfontsize\miniscule{5}{6}}
\fi
\makeatother

\DeclareRobustCommand{\optbar}[1]{\shortstack{{\miniscule (\rule[.5ex]{1.25em}{.18mm})}
  \\ [-.7ex] $#1$}}

\def\epm        {{\ensuremath{\Pe^\pm}}\xspace}

\def\mup        {{\ensuremath{\Pmu^+}}\xspace}
\def\mun        {{\ensuremath{\Pmu^-}}\xspace} 
\def\mupm       {{\ensuremath{\Pmu^\pm}}\xspace}

\def\W      {{\ensuremath{\PW}}\xspace}
\def\Wp     {{\ensuremath{\PW^+}}\xspace}
\def\Wm     {{\ensuremath{\PW^-}}\xspace}
\def\Wpm    {{\ensuremath{\PW^\pm}}\xspace}
\def\Z      {{\ensuremath{\PZ}}\xspace}

\def\uquark    {{\ensuremath{\Pu}}\xspace}

\def\dquark    {{\ensuremath{\Pd}}\xspace}

\def\squark    {{\ensuremath{\Ps}}\xspace}

\def\cquark    {{\ensuremath{\Pc}}\xspace}
\def\cquarkbar {{\ensuremath{\overline \cquark}}\xspace}
\def\ccbar     {{\ensuremath{\cquark\cquarkbar}}\xspace}
\def\bquark    {{\ensuremath{\Pb}}\xspace}
\def\bquarkbar {{\ensuremath{\overline \bquark}}\xspace}
\def\bbbar     {{\ensuremath{\bquark\bquarkbar}}\xspace}
\def\tquark    {{\ensuremath{\Pt}}\xspace}
\def\tquarkbar {{\ensuremath{\overline \tquark}}\xspace}
\def\ttbar     {{\ensuremath{\tquark\tquarkbar}}\xspace}

\def\KorKbar {\kern \thebaroffset\optbar{\kern -\thebaroffset \PK}{}\xspace}

\def\D       {{\ensuremath{\PD}}\xspace}

\def\DorDbar {\kern \thebaroffset\optbar{\kern -\thebaroffset \PD}\xspace}

\def\Dp      {{\ensuremath{\D^+}}\xspace}
\def\Dm      {{\ensuremath{\D^-}}\xspace}

\def\DpDm    {\ensuremath{\Dp {\kern -0.16em \Dm}}\xspace}

\def\B       {{\ensuremath{\PB}}\xspace}

\def\BorBbar {\kern \thebaroffset\optbar{\kern -\thebaroffset \PB}\xspace}

\def\Bd      {{\ensuremath{\B^0}}\xspace}

\def\BdorBdbar {\kern \thebaroffset\optbar{\kern -\thebaroffset \Bd}\xspace}

\def\Bs      {{\ensuremath{\B^0_\squark}}\xspace}

\def\BsorBsbar {\kern \thebaroffset\optbar{\kern -\thebaroffset \Bs}\xspace}

\def\jpsi     {{\ensuremath{{\PJ\mskip -3mu/\mskip -2mu\Ppsi}}}\xspace}

\def\Y#1S{\ensuremath{\PUpsilon{(#1S)}}\xspace}
\def\OneS  {{\Y1S}\xspace}

\def\LorLbar     {\kern \thebaroffset\optbar{\kern -\thebaroffset \PLambda}\xspace}

\newcommand{\decay}[2]{\ensuremath{\mathinner{#1\!\to #2}}\xspace}

\def\to                 {\ensuremath{\rightarrow}\xspace}

\def\AT#1     {\ensuremath{A_{\mathrm{T}}^{#1}}\xspace}           

\def\C#1      {\ensuremath{\mathcal{C}_{#1}}\xspace}                       
\def\Cp#1     {\ensuremath{\mathcal{C}_{#1}^{'}}\xspace}                    
\def\Ceff#1   {\ensuremath{\mathcal{C}_{#1}^{\mathrm{(eff)}}}\xspace}        
\def\Cpeff#1  {\ensuremath{\mathcal{C}_{#1}^{'\mathrm{(eff)}}}\xspace}       
\def\Ope#1    {\ensuremath{\mathcal{O}_{#1}}\xspace}                       
\def\Opep#1   {\ensuremath{\mathcal{O}_{#1}^{'}}\xspace}                    

\newcommand{\aunit}[1]{\ensuremath{\text{\,#1}}}       

\newcommand{\tev}{\aunit{Te\kern -0.1em V}\xspace}
\newcommand{\gev}{\aunit{Ge\kern -0.1em V}\xspace}
\newcommand{\mev}{\aunit{Me\kern -0.1em V}\xspace}
\newcommand{\kev}{\aunit{ke\kern -0.1em V}\xspace}
\newcommand{\ev}{\aunit{e\kern -0.1em V}\xspace}
 
\newcommand{\mevc}{\ensuremath{\aunit{Me\kern -0.1em V\!/}c}\xspace}
\newcommand{\gevc}{\ensuremath{\aunit{Ge\kern -0.1em V\!/}c}\xspace}
\newcommand{\mevcc}{\ensuremath{\aunit{Me\kern -0.1em V\!/}c^2}\xspace}
\newcommand{\gevcc}{\ensuremath{\aunit{Ge\kern -0.1em V\!/}c^2}\xspace}

\def\fb   {\ensuremath{\aunit{fb}}\xspace}
\def\invfb   {\ensuremath{\fb^{-1}}\xspace}

\def\deriv {\ensuremath{\mathrm{d}}}

\def\gsim{{~\raise.15em\hbox{$>$}\kern-.85em
          \lower.35em\hbox{$\sim$}~}\xspace}
\def\lsim{{~\raise.15em\hbox{$<$}\kern-.85em
          \lower.35em\hbox{$\sim$}~}\xspace}

\def\sqs   {\ensuremath{\protect\sqrt{s}}\xspace}

\def\pt         {\ensuremath{p_{\mathrm{T}}}\xspace}

\def\geant      {\mbox{\textsc{Geant4}}\xspace}

\def\photos     {\mbox{\textsc{Photos}}\xspace}
\def\powheg     {\mbox{\textsc{Powheg}}\xspace}
\def\pythia     {\mbox{\textsc{Pythia}}\xspace}
\def\resbos     {\mbox{\textsc{ResBos}}\xspace}

\def\tell1  {TELL1\xspace}
\def\ukl1   {UKL1\xspace}

\newcommand{\phz}{\phantom{0}}
\newcommand{\lhcborcid}[1]{\href{https://orcid.org/#1}{\hspace*{0.1em}\raisebox{-0.45ex}{\includegraphics[width=1em]{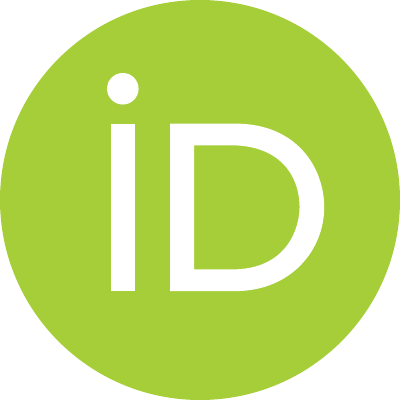}}}}

\hypersetup{
  colorlinks   = true, 
  urlcolor     = blue, 
  linkcolor    = blue, 
  citecolor    = red   
}

\ifEnableSectionTOCLinks
    \usepackage[explicit]{titlesec} 
    
    \let\oldcontentsline\contentsline
    \renewcommand

    \titleformat{\section}{\normalfont\Large\bf}{\hyperlink{tocsection.\thesection}{{\thesection} \parbox[t]{\dimexpr\textwidth-1pc}{#1}}}{1pc}{}

    \titleformat{\subsection}{\normalfont\bf}{\hyperlink{tocsubsection.\thesubsection}{{\thesubsection} \parbox[t]{\dimexpr\textwidth-1pc}{#1}}}{1pc}{}

    \titleformat{name=\section,numberless}[display]{}{}{0pt}{\normalfont\Huge\bfseries #1}
\fi

\usepackage{cite} 
\usepackage{LHCb/mciteplus}

\usepackage{longtable} 
\usepackage{rotating}
\usepackage{siunitx}
\usepackage{hyperref}

\begin{document}

\renewcommand{\thefootnote}{\fnsymbol{footnote}}
\setcounter{footnote}{1}


\begin{titlepage}
\pagenumbering{roman}

\vspace*{-1.5cm}
\centerline{\large EUROPEAN ORGANIZATION FOR NUCLEAR RESEARCH (CERN)}
\vspace*{1.5cm}
\noindent
\begin{tabular*}{\linewidth}{lc@{\extracolsep{\fill}}r@{\extracolsep{0pt}}}
\ifthenelse{\boolean{pdflatex}}
{\vspace*{-1.5cm}\mbox{\!\!\!\includegraphics[width=.14\textwidth]{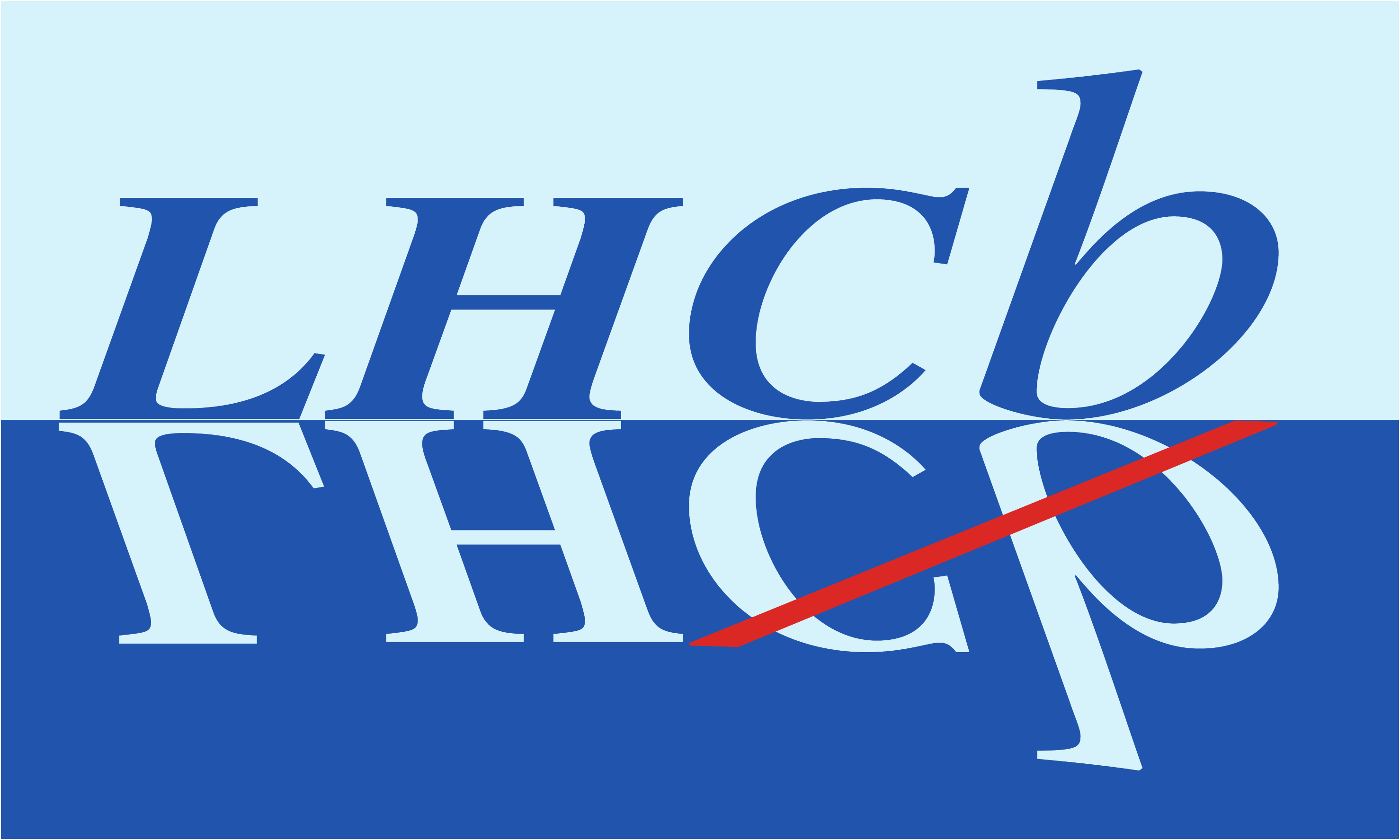}} & &}%
{\vspace*{-1.2cm}\mbox{\!\!\!\includegraphics[width=.12\textwidth]{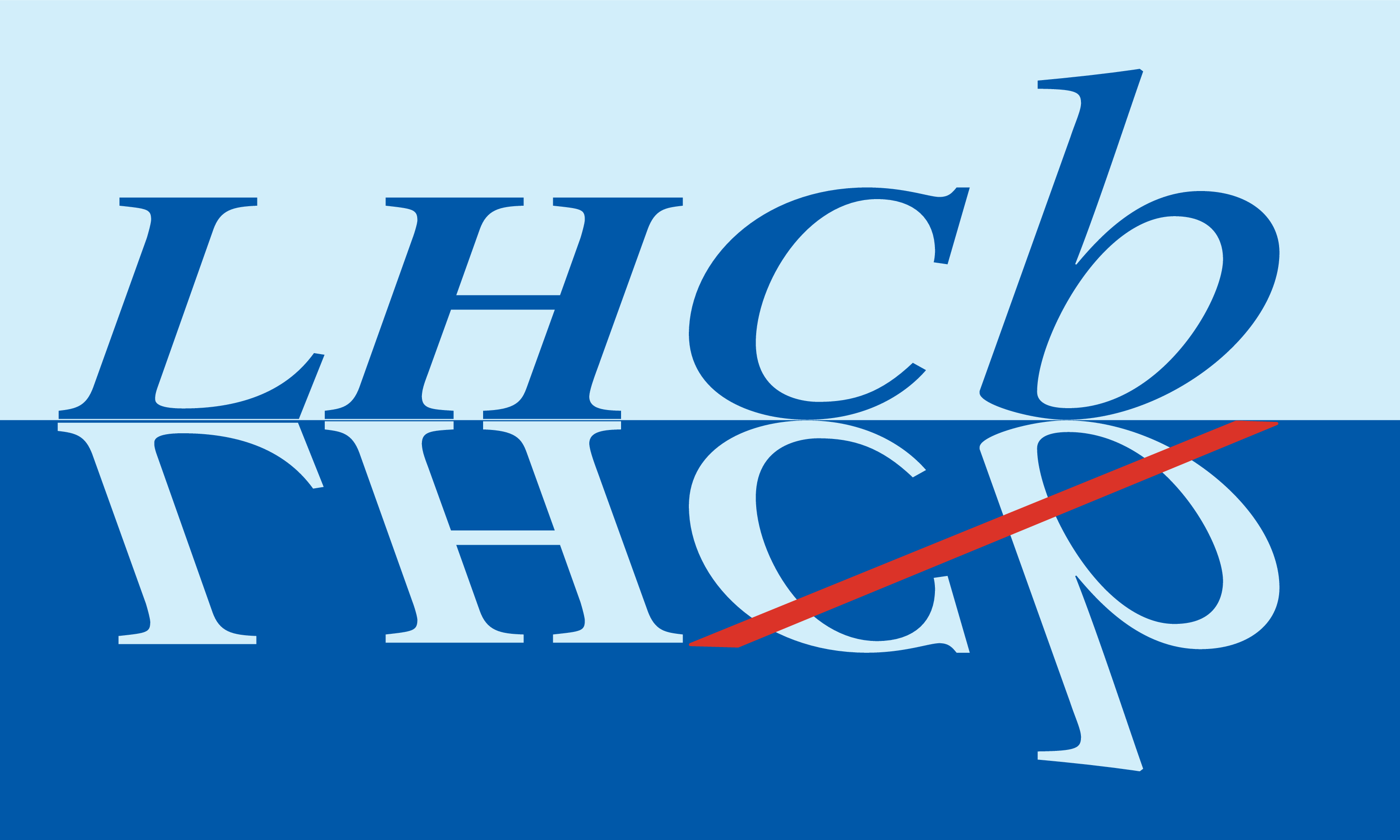}} & &}%
\\
 & & CERN-EP-2026-084 \\  
 & & LHCb-PAPER-2025-071 \\  
 & & \today \\ 
 & & \\
\end{tabular*}

\vspace*{4.0cm}

{\normalfont\bfseries\boldmath\huge
\begin{center}
  \papertitle 
\end{center}
}

\vspace*{2.0cm}

\begin{center}
\paperauthors\footnote{Authors are listed at the end of this Letter.}
\end{center}

\vspace{\fill}

\begin{abstract}
  \noindent
  A precision measurement of the muon charge asymmetry from 
$W$-boson decays in proton-proton collisions at \sqs = 13\tev is presented. The analysis utilizes data corresponding to an integrated luminosity of 5.1\invfb, recorded by the \lhcb detector during 2016, 2017 and 2018. The asymmetry is measured for muons with transverse momentum between 25 and 55\gev and pseudorapidity between 2.0 and 4.5. This result represents the most precise determination of the muon charge asymmetry in the forward region to date, exhibiting excellent agreement with next-to-next-to-leading-order predictions in perturbative quantum chromodynamics.
\end{abstract}

\vspace*{2.0cm}

\begin{center}
  Published in Phys.~Rev.~Lett. 137, 111801 (2026)
\end{center}

\vspace{\fill}

{\footnotesize 
\centerline{\copyright~\papercopyright. \href{\paperlicenceurl}{\paperlicence}.}}
\vspace*{2mm}

\end{titlepage}


\newpage
\setcounter{page}{2}
\mbox{~}
%
%
%
%


\renewcommand{\thefootnote}{\arabic{footnote}}
\setcounter{footnote}{0}


\cleardoublepage


\pagestyle{plain} 
\setcounter{page}{1}
\pagenumbering{arabic}



Precision measurements of electroweak boson production in proton-proton ($pp$) collisions~\cite{LHCb-PAPER-2012-008,LHCb-PAPER-2012-036,LHCb-PAPER-2014-033,LHCb-PAPER-2015-003,LHCb-PAPER-2015-001,LHCb-PAPER-2015-049,LHCb-PAPER-2016-021,LHCb-PAPER-2016-024,LHCb-PAPER-2021-037,LHCb-PAPER-2023-010} provide crucial constraints on the parton distribution functions (PDFs) of the proton~\cite{PhysRevD.93.033006, Ball2015, Harland-Lang:2014zoa, PhysRevD.89.054028}, especially for the light-flavor (\uquark, \dquark and \squark) quark and antiquark components. The lepton charge asymmetry in \decay{W}{\ell\nu} decays, where $\ell$ denotes the charged lepton (\epm, \mupm) from the \W-boson decay, is particularly sensitive to PDFs of valence quarks, is essential in global PDF analyses, and has been extensively studied at both the Tevatron~\cite{CDF:1998uzn,CDF:2005cgc,D0:2007pcy,D0:2008cgv,D0:2013xqc,D0:2014kma,CDF:2009cjw,D0:2013lql} and the Large Hadron Collider (\lhc)~\cite{LHCb-PAPER-2014-033,LHCb-PAPER-2015-049,LHCb-PAPER-2016-024,ATLAS:2011pph,ATLAS:2016nqi,CMS:2011bet,CMS:2012ivw,CMS:2013pzl,CMS:2016qqr,ATLAS:2025ede,ATLAS:2018pyl,ATLAS:2019fgb}.

At the \lhc, the proton's composition, comprising two valence \uquark quarks and one valence \dquark quark, leads to a higher production rate of \Wp than \Wm bosons. This imbalance results in a measurable asymmetry, $\cal{A}$, in the pseudorapidity distribution of the final-state charged leptons defined as
\begin{equation}
\label{eq:asym}
    {\cal{A}(\eta_\mu)} \equiv \frac{\deriv\sigma_{\Wp \to \mup \nu} /\deriv\eta _{\mup} -\deriv\sigma_{\Wm \to \mun \overline{\nu}}/\deriv\eta _{\mun}}{\deriv\sigma_{\Wp \to \mup \nu} /\deriv\eta _{\mup} + \deriv\sigma_{\Wm \to \mun \overline{\nu}}/\deriv\eta _{\mun}},
\end{equation}
where $\eta_\mu$ is the muon pseudorapidity in the laboratory frame and $\deriv\sigma/\deriv\eta_\mu $ is the differential cross-section for \decay{W}{\mu\nu} decays.
Measurements in the forward region provide unique sensitivity to PDFs with both small and large Bjorken-$x$, as needed for the interpretation of \lhc precision measurements~\cite{LHCb-PAPER-2021-024,LHCb-PAPER-2024-028,LHCb-PAPER-2021-048} and searches for new physics~\cite{Beenakker:2015rna}.
The muon charge asymmetry measurements at \lhcb particularly constrain the quark PDFs in the Bjorken-$x$ range $10^{-4} < x < 10^{-1}$~\cite{Deng_2021}, complementing \atlas and \cms results. 

This Letter presents results with significantly improved precision for the muon charge asymmetry in inclusive $pp \to W (\to \mu X) X^\prime$ production at \sqs = 13\tev, where $X^{(\prime)}$ represents additional decay products not used in the measurement. The analysis utilizes a data sample corresponding to an integrated luminosity of 5.1\invfb, collected with the \lhcb detector during 2016--2018. A more comprehensive description of the associated $W$-boson production cross-section measurement is reported in a companion article~\cite{LHCb:2026dan}.

The LHCb detector is a single-arm forward spectrometer covering the pseudorapidity range $2 < \eta < 5$, described in detail in Refs.~\cite{LHCb-DP-2008-001,LHCb-DP-2014-002}. Simulation is required to model the effects of the detector acceptance and the imposed selection requirements, and to describe the signal and background components when fitting data to extract yields. In the simulation, $pp$ collisions are generated using \pythia~\cite{Sjostrand:2007gs,*Sjostrand:2006za} with a specific \lhcb configuration~\cite{LHCb-PROC-2010-056}. 
 The interaction of the generated particles with the detector, and its response, are implemented using the \geant toolkit~\cite{Allison:2006ve, *Agostinelli:2002hh} as described in Ref.~\cite{LHCb-PROC-2011-006}. 

 To account for higher-order quantum chromodynamics (QCD) effects, the leading-order \pythia predictions are weighted to next-to-next-to-leading-order calculations from \textsc{DYTurbo}~\cite{DYTurbo2020}.
Both angular distribution and unpolarized cross-section correction factors are employed.
Angular coefficients are determined through a projection method~\cite{Gauld:2017tww} that computes normalized weighted averages over the dilepton angular coordinates ($\theta$, $\phi$). 
For the unpolarized cross-section corrections, simulated events are weighted using interpolation histograms. This involves trilinear interpolation between histogram bins, with the histogram arrays typically parameterized by quantities such as $\alpha_{\rm s}$ (the strong coupling constant) and $k_{\rm T}$ (the intrinsic transverse momentum of the initial state partons). 

Candidate \decay{W}{\mu\nu} decays are reconstructed from muons with momentum $p^{\mu} <2\tev$ and satisfying the fiducial requirements of transverse momentum $25< \pt^{\mu} <55\gev$ and $2.0< \eta_\mu <4.5$ (natural units with $\hbar=c=1$ are used throughout). The selected muons are required to have a relative momentum uncertainty of less than 6\%, to be consistent with originating from a \textit{pp} primary collision vertex, and to have good track-fit quality~\cite{LHCb-DP-2014-001, LHCb-DP-2017-001}.
Furthermore, the muon isolation, defined as the scalar sum of the transverse momenta of all other particles within $\sqrt{(\phi - \phi_\mu)^2 + (\eta - \eta_\mu)^2} < 0.5$, where $\phi$ is the azimuthal angle, is required to be less than 3\gev. The analysis employs a \Z-boson veto to exclude events with secondary muons that might come from \decay{Z}{\mu\mu} decays.
After all selection criteria, about $6.3 \times 10^6$ ($4.4 \times 10^6$) \Wp (\Wm) candidates are retained with an approximate purity of 85\% (80\%).

The muon charge asymmetry, $\cal{A}$, is measured via Eq.~\ref{eq:asym}, in which the cross-section 
$\sigma_{\Wpm \to \mupm \nu} = \frac{N^{\pm} \cdot f^{W^{\pm}}_{\rm FSR}}{ \mathcal{L} \cdot \varepsilon_{W^{\pm}}}$ 
is evaluated for each $\eta_\mu$ interval, where $\mathcal{L}$ denotes the integrated luminosity and $N^{\pm}$ the measured \Wpm candidate yield.
The total efficiency $\varepsilon_{W^{\pm}}$ is determined as a fraction from simulation, with the numerator comprising weighted \W-boson events passing all selection criteria and the denominator containing generator-level events satisfying only fiducial requirements. Correction weights are applied to the simulation and
account for tracking, muon-identification, triggering, and selection efficiencies. The factor $f^{\Wpm}_{\rm FSR}$ represents the final-state radiation (FSR) correction determined using \photos~\cite{davidson2015photos}.
In the muon charge asymmetry measurement, the integrated luminosity cancels out, eliminating the need for associated systematic uncertainty assignments.
Given the excellent angular resolution of the \lhcb detector, migration between $\eta_{\mu}$ intervals is negligible.

To suppress the effects of detector misalignment, charge-dependent curvature biases are corrected through the pseudomass method~\cite{Barter:2021npd,LHCb-DP-2023-001} using \decay{Z}{\mu\mu} decays in data. 
A simultaneous fit to the invariant masses of $\jpsi$, \OneS and \Z-boson dimuon final-states provides constraints on the smearing parameters of muon momentum corrections in simulation. 

The background contributions from heavy-flavor ($\bbbar$, $\ccbar$, $\ttbar$) and electroweak processes ($W\to\tau\nu$, \decay{Z}{\mu\mu}, $Z\to\tau\tau$, $WW$, Drell--Yan and \tquark-quark production) are estimated using simulated samples. These are calibrated and corrected with the selected \decay{Z}{\mu\mu} data candidates across $\eta_\mu$ intervals.
The dominant background contribution originates from hadrons that are misidentified as muons. The ratio of the $\pt^{\mu}$ distributions between data and simulation and the misidentification efficiencies are parameterized and applied as event weights in the simulation to better match the data.

\begin{figure}[!tbp]
  \includegraphics[width=0.49\textwidth] {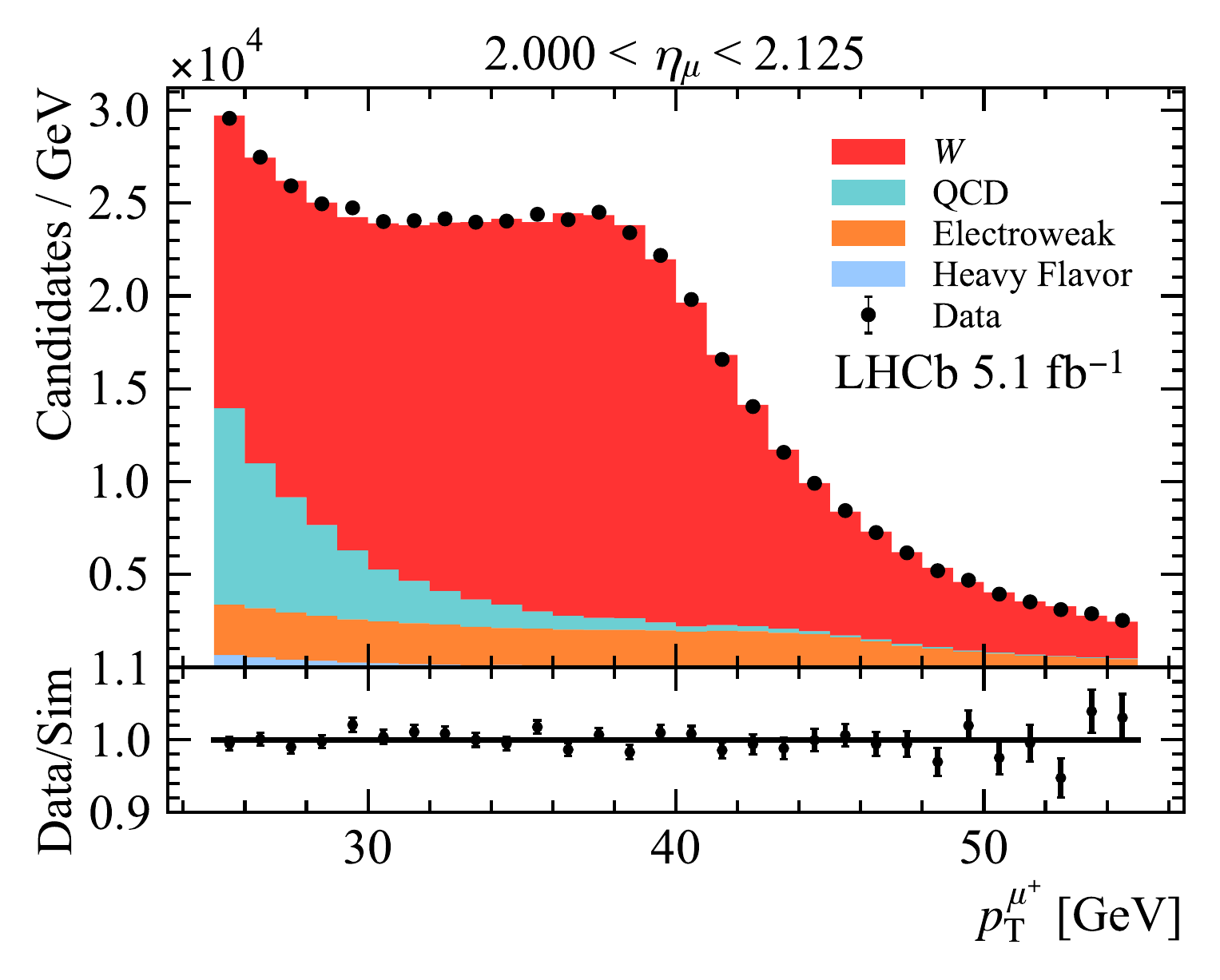}
  \includegraphics[width=0.49\textwidth] {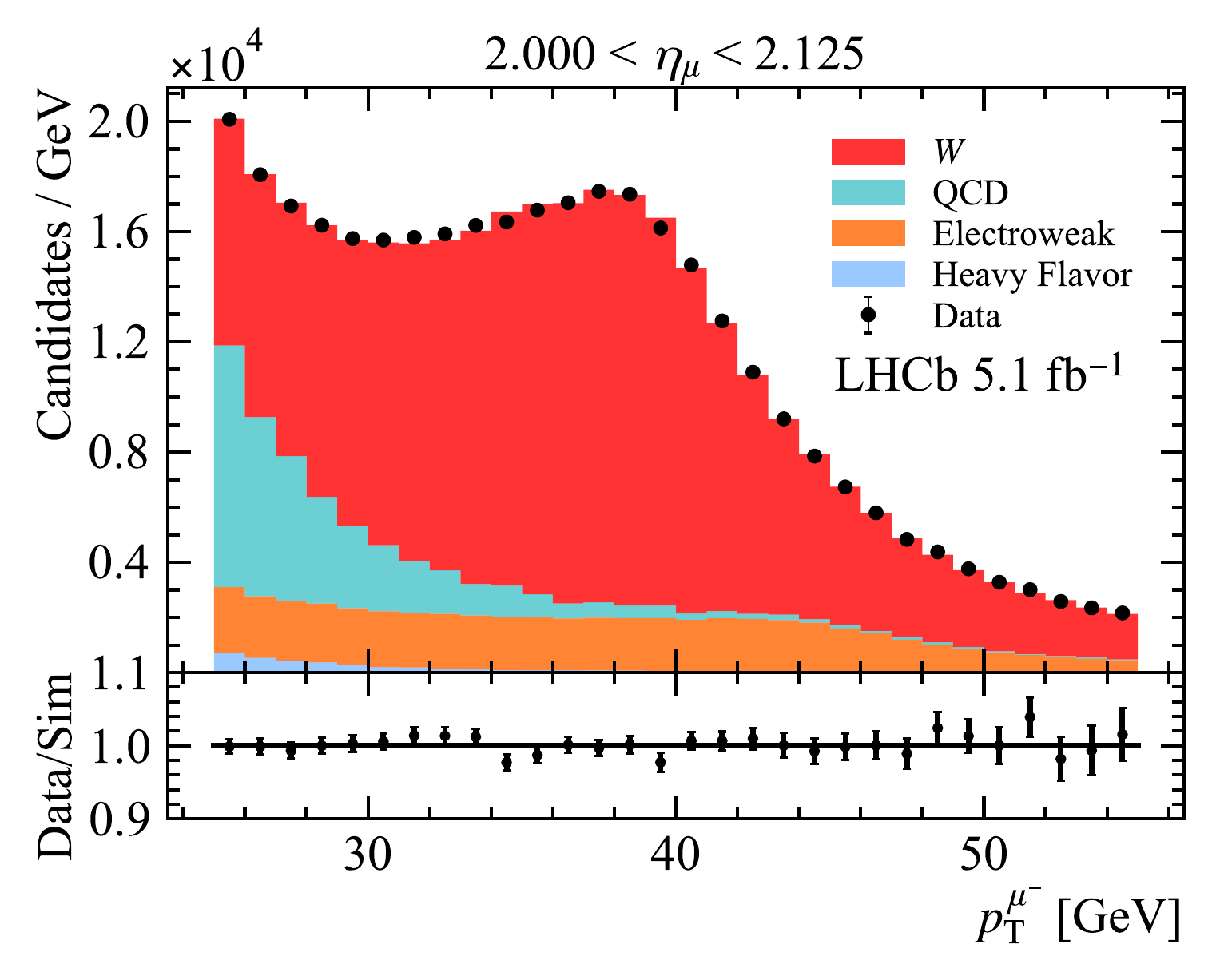}
  \caption{Example $\pt^\mu$ distributions in data for (left) \mup  and (right) \mun, with the results of the fit also shown. Error bars represent statistical uncertainties only.} \label{fig:fit}
\end{figure}

The signal extraction employs a separate template-based binned likelihood fit for each \W-boson charge in the 18 $\eta_\mu$ intervals within the fiducial region.
In the fitting procedure, both \WToMuNu signal and QCD background fractions are treated as free parameters, fixing electroweak and heavy-flavor fractions to theory predictions. Statistical uncertainties in the template modeling are addressed through the Beeston--Barlow lite prescription~\cite{BeestonBarlow:1993}.
The signal fraction in the data ranges from 63\% to 88\%, depending on the muon charge and $\eta_\mu$ region. 
Figure~\ref{fig:fit} displays examples of the signal extraction results for the first pseudorapidity bin, $2.000 < \eta_\mu < 2.125$. 

The muon reconstruction efficiency is determined using a tag-and-probe method~\cite{LHCb-DP-2013-002} in $\eta_\mu$ intervals from \decay{Z}{\mu\mu} and \decay{\OneS}{\mu\mu} samples~\cite{LHCb-PAPER-2021-024}.
The efficiency components are factorized into three contributions: the tracking efficiency quantifies the reconstruction probability of a charged track within the LHCb fiducial acceptance; the muon identification efficiency measures the fraction of tracks consistent with the muon hypothesis; and the trigger efficiency represents the fraction of identified muons passing the hardware trigger selection. 
To account for kinematic dependencies, the efficiency corrections are implemented with a binned approach across the 18 $\eta_\mu$ intervals, with each interval featuring an independent $\pt^{\mu}$ parameterization. For all three terms, the ratios of efficiencies for data and simulation are used to weight simulated muon candidates. For all other selection requirements, a two-dimensional binning scheme in $u$ and $\eta_\mu$ is used instead, where the recoil projection variable $u = \frac{\vec{p}_{\rm T}^{\,V} \cdot \vec{p}_{\rm T}^{\,\mu}} {\pt^{\mu}}$ quantifies the \pt alignment between the muon and the parent boson ($V$).

Several systematic uncertainties affecting the measured muon charge asymmetry are considered, as listed in Table~\ref{tab:asy_syst}. 
The reconstruction efficiency uncertainties comprise the limited tag-and-probe sample size, the fit-function dependence, and the kinematic adjustments to $\pt^{\mu}$ thresholds and \Z-boson mass-window definitions. 
Uncertainties from generator modeling include scale variations in \textsc{DYTurbo}, comparisons between \textsc{DYTurbo} and \powheg predictions, and PDF variations using  MSHT20~\cite{Bailey:2020ooq}, NNPDF31~\cite{NNPDF:2017mvq}, and CT18~\cite{Hou:2019efy} PDFs sets. 
Uncertainties from the smearing model used to correct the simulated muon momenta are propagated in the fit.
Selection efficiency uncertainties arise from limited data and simulation sample size, binning-scheme-dependent event topology variations, and the Z-boson selection criteria definitions. Uncertainties from QCD background modeling arise from the fluctuations in the ratio of data to simulation samples, misidentification function variations, and hadronic composition differences among pions, kaons, and protons. The FSR uncertainties are estimated by comparing differences of corrections between \resbos + \photos and \powheg + \pythia.

\begin{table}[!tbp]
  \centering
  \caption{\label{tab:asy_syst} Summary of uncertainty contributions (in units of $10^{-3}$) on the $W$-boson charge asymmetry from the reconstruction efficiency ($\varepsilon_{\rm rec}$), generator correction, smearing, selection efficiency ($\varepsilon_{\rm sel}$), QCD background and FSR correction to the measured results.}
\begin{tabular}{c|cccccc}
\hline
$\eta_{\mu}$   & $\varepsilon_{\rm rec}$   & Generator    &Smearing    &$\varepsilon_{\rm sel}$   &QCD Bkg   &FSR \\
\hline
2.000 -- 2.125  & 3.6 & 2.6 & 0.8 & 0.2 & 0.6 & 0.3\\
2.125 -- 2.250  & 2.7 & 3.4 & 0.4 & 0.2 & 0.3 & 1.0\\
2.250 -- 2.375  & 2.7 & 3.2 & 0.5 & 0.2 & 0.3 & 0.2\\
2.375 -- 2.500  & 2.7 & 3.8 & 0.3 & 0.2 & 0.3 & 0.7\\
2.500 -- 2.625  & 1.9 & 3.8 & 0.4 & 0.2 & 0.3 & 0.6\\
2.625 -- 2.750  & 1.9 & 3.3 & 0.4 & 0.2 & 0.2 & 0.6\\
2.750 -- 2.875  & 2.6 & 2.3 & 0.5 & 0.1 & 0.5 & 0.1\\
2.875 -- 3.000  & 2.7 & 2.3 & 0.3 & 0.1 & 0.5 & 0.4\\
3.000 -- 3.125  & 3.1 & 1.8 & 0.5 & 0.1 & 0.3 & 0.1\\
3.125 -- 3.250  & 3.3 & 1.4 & 0.4 & 0.1 & 0.7 & 0.6\\
3.250 -- 3.375  & 2.8 & 3.0 & 0.7 & 0.1 & 0.8 & 0.6\\
3.375 -- 3.500  & 2.6 & 6.5 & 0.5 & 0.1 & 1.0 & 0.3\\
3.500 -- 3.625  & 3.8 & 9.5 & 0.6 & 0.1 & 0.4 & 0.5\\
3.625 -- 3.750  & 3.3 & 11.0 & 0.8 & 0.2 & 0.4 & 0.1\\
3.750 -- 3.875  & 3.1 & 13.2 & 1.1 & 0.1 & 0.4 & 2.2\\
3.875 -- 4.000  & 3.8 & 14.3 & 0.6 & 0.2 & 0.4 & 0.1\\
4.000 -- 4.250  & 3.5 & 19.4 & 0.8 & 1.0 & 0.7 & 0.4\\
4.250 -- 4.500  & 6.1 & 13.7 & 2.4 & 1.2 & 1.5 & 1.4\\
\hline
\end{tabular}
\end{table}

\begin{figure}[!htbp]
  \begin{center}                     
  \includegraphics[width=0.8\textwidth] {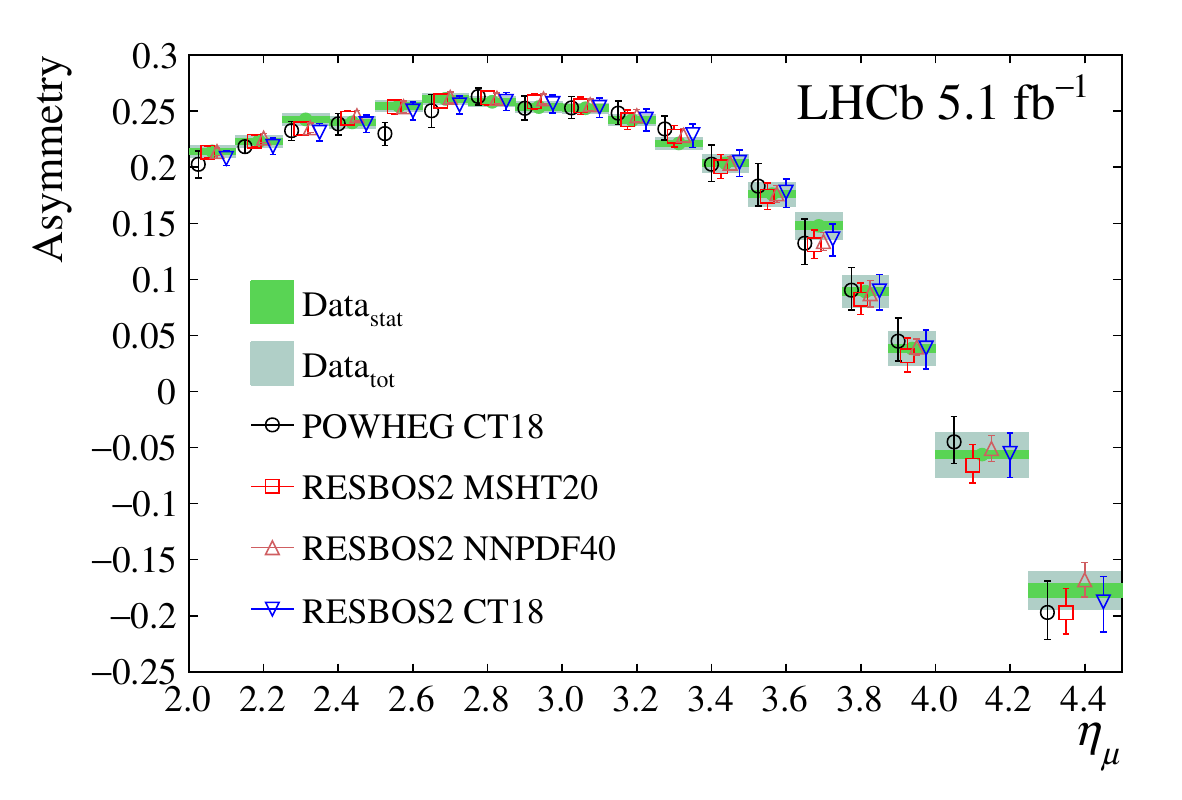}
  \end{center}
  \caption{Measured muon charge asymmetry, compared to the predictions with different generators and PDF sets. The bands correspond to the data, with the inner (outer) region corresponding to the statistical (total) uncertainty. The uncertainty bars of the predictions correspond to a 68\% confidence level.} \label{fig:asy}
\end{figure}

Figure~\ref{fig:asy} and Table~\ref{tab:data_theory_asy} present the asymmetry measurement.
For the measured results, both statistical and systematic uncertainties are presented, compared with theoretical predictions from \powheg (using CT18) and \resbos2~\cite{PhysRevD.86.094026} (using MSHT20, NNPDF40~\cite{NNPDF:2021njg}, and CT18 PDF sets). Theoretical uncertainties arise from PDF uncertainties~\cite{bourilkov2006} and scale-dependent effects. The measured results show good agreement with theoretical predictions.
Across all $\eta_\mu$ intervals, the experimental uncertainties are comparable in magnitude to the PDF uncertainties from the MSHT20, NNPDF40, and CT18 sets and the scale uncertainties. This demonstrates the significance of the current analysis for constraining and improving PDFs.

\begin{table}[!tbp]
  \centering
  \caption{\label{tab:data_theory_asy} Measured muon charge asymmetries $\cal{A}(\eta_\mu)$ (in units of $10^{-3}$) with (first) statistical and (second) systematic uncertainties. Theoretical predictions are indicated for \powheg with the CT18 PDF set and \resbos2 with three PDF sets. Their uncertainties arise from PDF and scale variations.}
\begin{tabular}{crrrrr}
\hline
$\eta_\mu$   & Measurement   & \multicolumn{1}{c}{\powheg}   & \multicolumn{1}{c}{MSHT20}   & \multicolumn{1}{c}{NNPDF40}   & \multicolumn{1}{c}{CT18} \\
\hline
2.000 -- 2.125    & $214\pm3\pm5\phz$      & $203^{\:+\:12}_{\:-\:12}$      & $213^{\:+\:5\phz}_{\:-\:5\phz}$      & $214\pm3\phz$      & $208^{\:+\:7\phz}_{\:-\:6\phz}$ \\
2.125 -- 2.250    & $223\pm3\pm4\phz$      & $218^{\:+\:5\phz}_{\:-\:6\phz}$        & $223^{\:+\:6\phz}_{\:-\:5\phz}$        & $226\pm4\phz$        & $218^{\:+\:7\phz}_{\:-\:7\phz}$ \\
2.250 -- 2.375    & $243\pm3\pm4\phz$      & $233^{\:+\:8\phz}_{\:-\:9\phz}$        & $234^{\:+\:6\phz}_{\:-\:6\phz}$        & $235\pm4\phz$        & $231^{\:+\:8\phz}_{\:-\:8\phz}$ \\
2.375 -- 2.500    & $240\pm3\pm5\phz$      & $239^{\:+\:9\phz}_{\:-\:10}$       & $244^{\:+\:6\phz}_{\:-\:6\phz}$       & $246\pm5\phz$       & $239^{\:+\:8\phz}_{\:-\:8\phz}$ \\
2.500 -- 2.625    & $255\pm3\pm4\phz$      & $230^{\:+\:10}_{\:-\:11}$      & $254^{\:+\:7\phz}_{\:-\:7\phz}$      & $254\pm5\phz$      & $250^{\:+\:8\phz}_{\:-\:8\phz}$ \\
2.625 -- 2.750    & $261\pm3\pm4\phz$      & $250^{\:+\:14}_{\:-\:15}$      & $259^{\:+\:7\phz}_{\:-\:7\phz}$      & $262\pm5\phz$      & $256^{\:+\:8\phz}_{\:-\:8\phz}$ \\
2.750 -- 2.875    & $258\pm3\pm4\phz$      & $263^{\:+\:7\phz}_{\:-\:8\phz}$        & $262^{\:+\:7\phz}_{\:-\:7\phz}$        & $261\pm5\phz$        & $259^{\:+\:8\phz}_{\:-\:8\phz}$ \\
2.875 -- 3.000    & $254\pm3\pm4\phz$      & $253^{\:+\:11}_{\:-\:11}$      & $258^{\:+\:7\phz}_{\:-\:7\phz}$      & $261\pm4\phz$      & $257^{\:+\:8\phz}_{\:-\:8\phz}$ \\
3.000 -- 3.125    & $253\pm3\pm4\phz$      & $253^{\:+\:10}_{\:-\:10}$      & $255^{\:+\:8\phz}_{\:-\:8\phz}$      & $256\pm5\phz$      & $253^{\:+\:8\phz}_{\:-\:9\phz}$ \\
3.125 -- 3.250    & $242\pm3\pm4\phz$      & $248^{\:+\:11}_{\:-\:10}$      & $242^{\:+\:9\phz}_{\:-\:9\phz}$      & $246\pm5\phz$      & $243^{\:+\:9\phz}_{\:-\:10}$ \\
3.250 -- 3.375    & $221\pm3\pm4\phz$      & $234^{\:+\:11}_{\:-\:10}$      & $228^{\:+\:10}_{\:-\:10}$    & $228\pm6\phz$    & $229^{\:+\:10}_{\:-\:11}$ \\
3.375 -- 3.500    & $204\pm3\pm7\phz$      & $203^{\:+\:17}_{\:-\:15}$      & $200^{\:+\:11}_{\:-\:11}$    & $204\pm6\phz$    & $204^{\:+\:11}_{\:-\:13}$ \\
3.500 -- 3.625    & $176\pm3\pm10$      & $183^{\:+\:20}_{\:-\:18}$     & $174^{\:+\:12}_{\:-\:11}$   & $176\pm8\phz$   & $178^{\:+\:12}_{\:-\:14}$ \\
3.625 -- 3.750    & $148\pm4\pm12$      & $132^{\:+\:21}_{\:-\:19}$     & $131^{\:+\:13}_{\:-\:12}$   & $134\pm8\phz$   & $136^{\:+\:13}_{\:-\:15}$ \\
3.750 -- 3.875    & $89\pm4\pm14$      & $90^{\:+\:20}_{\:-\:18}$       & $82^{\:+\:15}_{\:-\:13}$      & $87\pm12$      & $90^{\:+\:14}_{\:-\:17}$ \\
3.875 -- 4.000    & $38\pm4\pm15$      & $45^{\:+\:21}_{\:-\:18}$       & $32^{\:+\:16}_{\:-\:14}$      & $40\pm7\phz$       & $39^{\:+\:16}_{\:-\:19}$ \\
4.000 -- 4.250    & $-56\pm3\pm20$      & $-45^{\:+\:23}_{\:-\:19}$     & $-66^{\:+\:18}_{\:-\:16}$      & $-51\pm11$      & $-55^{\:+\:18}_{\:-\:22}$ \\
4.250 -- 4.500    & $-177\pm6\pm15$      & $-197^{\:+\:28}_{\:-\:24}$   & $-197^{\:+\:22}_{\:-\:19}$      & $-168\pm15$      & $-188^{\:+\:22}_{\:-\:27}$ \\
\hline
\end{tabular}
\end{table}

In summary, a high-precision measurement is reported of the \Wpm-boson muon charge asymmetry in $pp$ collisions at $\sqs=13\tev$, using data collected by the \lhcb experiment during 2016--2018, corresponding to an integrated luminosity of 5.1\invfb. The asymmetry is measured for muons with transverse momentum $25 <\pt^{\mu}<55\gev$ and pseudorapidity $2.0 <\eta_{\mu}<4.5$, employing a binning scheme with finer granularity compared to previous \lhcb results, representing the most precise determination in the forward region to date.
The results show good agreement with next-to-next-to-leading-order QCD predictions. 
The precision measurement provides crucial experimental constraints for the global fit of proton PDFs.
The complete covariance matrix of the measured muon charge asymmetry and FSR corrections are provided in the End Matter, offering valuable input for future PDF determinations.

\section*{Acknowledgements}
%
%
\noindent We express our gratitude to our colleagues in the CERN
accelerator departments for the excellent performance of the LHC. We
thank the technical and administrative staff at the LHCb
institutes.
We acknowledge support from CERN and from the national agencies:
ARC (Australia);
CAPES, CNPq, FAPERJ and FINEP (Brazil); 
MOST and NSFC (China); 
CNRS/IN2P3 and CEA (France);  
BMFTR, DFG and MPG (Germany);
INFN (Italy); 
NWO (Netherlands); 
MNiSW and NCN (Poland); 
MEC/IFA (Romania); 
MICIU and AEI (Spain);
SNSF and SER (Switzerland); 
NASU (Ukraine); 
STFC (United Kingdom); 
DOE NP and NSF (USA).
We acknowledge the computing resources that are provided by ARDC (Australia), 
CBPF (Brazil),
CERN, 
IHEP and LZU (China),
IN2P3 (France), 
KIT and DESY (Germany), 
INFN (Italy), 
SURF (Netherlands),
Polish WLCG (Poland),
IFIN-HH (Romania), 
PIC (Spain), CSCS (Switzerland), 
GridPP (United Kingdom),
and NSF (USA).  
We are indebted to the communities behind the multiple open-source
software packages on which we depend.
Individual groups or members have received support from
RTP (Australia), 
Key Research Program of Frontier Sciences of CAS, CAS PIFI, CAS CCEPP (China); 
Minciencias (Colombia);
EPLANET, Marie Sk\l{}odowska-Curie Actions, ERC and NextGenerationEU (European Union);
A*MIDEX, ANR, IPhU and Labex P2IO, and R\'{e}gion Auvergne-Rh\^{o}ne-Alpes (France);
Alexander-von-Humboldt Foundation (Germany);
ICSC (Italy); 
Severo Ochoa and Mar\'ia de Maeztu Units of Excellence, GVA, XuntaGal, GENCAT, InTalent-Inditex and Prog.~Atracci\'on Talento CM (Spain);
the Leverhulme Trust, the Royal Society and UKRI (United Kingdom).

\clearpage


\clearpage
\addcontentsline{toc}{section}{References}
\bibliographystyle{LHCb/LHCb}
\bibliography{main,LHCb/standard,LHCb/LHCb-PAPER,LHCb/LHCb-CONF,LHCb/LHCb-DP,LHCb/LHCb-TDR,LHCb/LHCb-PUB}

\newpage
\centerline
{\large\bf LHCb collaboration}
\begin
{flushleft}
\small
R.~Aaij$^{38}$\lhcborcid{0000-0003-0533-1952},
M. ~Abdelfatah$^{69}$,
A.S.W.~Abdelmotteleb$^{57}$\lhcborcid{0000-0001-7905-0542},
C.~Abellan~Beteta$^{51}$\lhcborcid{0009-0009-0869-6798},
F.~Abudin{\'e}n$^{59}$\lhcborcid{0000-0002-6737-3528},
T.~Ackernley$^{61}$\lhcborcid{0000-0002-5951-3498},
A. A. ~Adefisoye$^{69}$\lhcborcid{0000-0003-2448-1550},
B.~Adeva$^{47}$\lhcborcid{0000-0001-9756-3712},
M.~Adinolfi$^{55}$\lhcborcid{0000-0002-1326-1264},
P.~Adlarson$^{87}$\lhcborcid{0000-0001-6280-3851},
C.~Agapopoulou$^{14}$\lhcborcid{0000-0002-2368-0147}
% ...

{\footnotesize \it

% ...
$^{14}$Universit{\'e} Paris-Saclay, CNRS/IN2P3, IJCLab, Orsay, France\\
% ...
$^{38}$Nikhef National Institute for Subatomic Physics, Amsterdam, Netherlands\\
% ...
$^{47}$Instituto Galego de F{\'\i}sica de Altas Enerx{\'\i}as (IGFAE), Universidade de Santiago de Compostela, Santiago de Compostela, Spain\\
% ...
$^{51}$Physik-Institut, Universit{\"a}t Z{\"u}rich, Z{\"u}rich, Switzerland\\
% ...
$^{55}$H.H. Wills Physics Laboratory, University of Bristol, Bristol, United Kingdom\\
% ...
$^{57}$Department of Physics, University of Warwick, Coventry, United Kingdom\\
% ...
$^{59}$School of Physics and Astronomy, University of Edinburgh, Edinburgh, United Kingdom\\
$^{60}$School of Physics and Astronomy, University of Glasgow, Glasgow, United Kingdom\\
$^{61}$Oliver Lodge Laboratory, University of Liverpool, Liverpool, United Kingdom\\
% ...
$^{69}$Syracuse University, Syracuse, NY, United States\\
% ...
$^{87}$Department of Physics and Astronomy, Uppsala University, Uppsala, Sweden, associated to $^{60}$\\
% ...
\medskip
}
\end{flushleft}

%
%
%
%

\end{document}